# Modeling and Optimization of Cascaded Four-Wave Mixing toward the VUV in Xenon-Filled Negative-Curvature Fibers

Jose Otavio Rosa,[1] Frédéric Gérôme,[2] Fetah Benabid,[2] Jonas H. Osório,[3] and Jonathas de Paula Siqueira[1,*]

[1]Gleb Wataghin Institute of Physics, University of Campinas – UNICAMP, 13083-859 Campinas, SP, Brazil
[2] GPPMM Group, XLIM Institute, CNRS UMR 7252, University of Limoges, Limoges, 87060, France
[3] Multiuser Laboratory of Optics and Photonics, LaMOF, Federal University of Lavras, Lavras, 37203-202, Brazil

*siquejp@unicamp.br

## Abstract

In this work, we thoroughly characterize third- and fourth-harmonic generation through cascaded four-wave mixing in a xenon-filled negative-curvature hollow-core fiber operated under a pressure gradient. The experimental results are supported by both a semi-analytical model and numerical simulations, which accurately reproduce the pressure dependence of the harmonic conversion. We identify the optimum pressure conditions for third and fourth-harmonic generation and experimentally observe the cascaded four-wave-mixing process extending into the vacuum ultraviolet (VUV), up to the sixth harmonic at 172 nm, with the accessible spectral range limited by the available VUV filter.

## Introduction

Ultrafast coherent vacuum ultraviolet (VUV: $\lambda < 200$ nm or $h\nu > 6$ eV) light sources are of increasing interest due to their wide range of applications such as photocatalysis, photoionization, and photoemission studies[1–3]. Specifically, photoemission-based spectroscopies and microscopies require photon energies higher than ~ 6 eV to overcome the work function of many materials[4]. In this context, a primary example of advanced spectroscopy employing laser-based VUV sources is Angle-Resolved Photoemission Spectroscopy (ARPES), where collecting photoemitted electrons, measuring both kinetic energy and emission angle, allows for the reconstruction of the material's band structure [5]. Furthermore, by employing ultrashort pulsed light sources in a pump-probe configuration, the experiment can be performed in its time-resolved version (tr-ARPES) [6]. This technique enables mapping the band structure above the Fermi level and the ultrafast photoexcited electron relaxation dynamics with high temporal, energy, and momentum resolution. As a result, it has played a critical role in unraveling the complex dynamical properties of quantum materials, including high $T_c$ superconductors [7] and topological insulators [8], among many other systems exhibiting strong electronic correlations investigated by tr-ARPES.

In this framework, optical materials for frequency conversion, like SBO ($SrB_4O_7$)[9] and KBBF ($KBe_2BO_3F_2$) [10], have been demonstrated to achieve the VUV region. However, ultraviolet generation in these crystals is limited due to the absorption around 130 nm for SBO and 147 nm for KBBF. In turn, another strategy for VUV generation is the use of structured crystals for quasi phase-matching, but it also has limitations for shorter wavelengths [11,12]. Additionally, these materials limit the generation of ultrashort pulses, due to the strong change of the refractive index near the absorption edge, limiting the phase-matching bandwidth [13].

To overcome the limitation of solid-state materials for frequency conversion, gaseous media have been employed to achieve wavelengths below 130 nm through perturbative processes, or even to the extreme ultraviolet (XUV) and X-ray region in the strong-field regime [14–16]. In the VUV region, these techniques are well-established for providing coherent, high-flux, and tunable light sources on a laboratory scale, offering performance comparable to that of large-scale, high-demand facilities such as synchrotrons and free-electron lasers (FELs) [17,18].

In particular, VUV generation via frequency conversion in gases is typically achieved by free-space focusing on gas jets or gas-filled chambers [19,20]. However, this configuration is limited by a short interaction length, characterized by the Rayleigh range, which inherently restricts generation efficiency. Another significant constraint is phase-matching; in free-focusing setups, this condition is often achieved when the harmonic wavelength lies near a gas resonance [21,22]. In such cases, anomalous dispersion is used to compensate for the contributions of free electrons and the Gouy phase shift. Consequently, this dependence restricts the source's tunability, a critical factor for specific photoemission experiments [23].

In the VUV region, as an alternative to harmonic generation through free-space focusing, phase matched four-wave mixing (FWM) in gas-filled hollow capillaries can be employed[24,25]. For the FWM process occurring between the fields coupled in the lowest order modes of a capillary waveguide, the waveguide dispersion can be balanced by the gas pressure, enabling phase-matching over a significantly longer interaction length as first demonstrated by Durfee et al. in a seminal work where the phase matched generation of energetic pulses at 260 nm were demonstrated [24,26]. In an extension of this work, Misoguti et al. demonstrated the efficient generation of VUV ultrashort pulses using a similar geometry through a cascaded FWM process [27]. While VUV regions can be reached through cascaded generation processes in these capillaries, the reliance on Fresnel reflections for confinement leads to high propagation losses for small core diameters, leading to the use of inner diameters higher than 100 $\mu$m. Given the high effective mode area of the eigenmodes of such capillaries, high energies (> 100 $\mu$J at pump frequencies) were necessary to achieve high conversion efficiencies, limiting the achievable repetition rate to a few kHz. To overcome these challenges, negative-curvature hollow-core fibers (NC-HCFs), also known in the literature as antiresonant hollow-core fibers, have emerged as a superior platform. Couch et al. have demonstrated highly efficient cascaded FWM in NC-HCFs reaching harmonics energies up to 18 eV using highly stable high repetition rate Yb based lasers[28,29]. Due to their unique guiding mechanism and high confinement, NC-HCFs provide low-loss propagation and enable efficient nonlinear conversion at high repetition rates [30]. This capability is particularly relevant for mitigating space-charge effects in photoemission experiments, where maintaining high average power with low pulse energy is essential [31].

In the seminal work of Couch et al. [30], it was shown through simulations that in cascaded harmonic generation processes, the first generated harmonic is of great importance for the subsequent generation of the higher order harmonics through the cascaded FWM mechanism. Consequently, optimization supported by modelling of the first harmonics is vital to improve its generation and maximize the yield of the cascaded FWM process up to higher VUV photon energies. However, a more detailed study concerning how the pressure gradient parameters affect this efficiency is lacking in the literature. In this work, we investigate the dependence of generation efficiency of the third and fourth harmonics of a femtosecond Yb laser (1030 nm) generated through FWM in NC-HCF filled with Xe gas under varying pressure gradient conditions. More specifically, we investigate how the efficiency of both harmonics is influenced by the initial pressure of the gradient starting from a given pressure and decreasing to vacuum conditions. We believe that the results reported herein provide new insights into the dynamics of cascaded FWM under pressure gradients, serving as a foundation for the development of high-repetition-rate ultrafast VUV sources.

## Experimental setup

In this work, the driving source for harmonic generation is a Yb:KGW regenerative CPA system (Carbide CB5-SP, Light Conversion Ltd) centered at 1030 nm with pulse durations of 170 fs operating at 200 kHz. The waveguide employed in this investigation is a tubular-lattice NC-HCF fiber with approximately 37 µm core diameter. Its cladding structure consists of eight untouching capillaries with a wall thickness of approximately 1.1 µm as shown in Figure 1-a). This NC-HCF has a low attenuation at the third harmonic of Yb:KGW fundamental wavelength (50 dB/km at 343 nm) used to drive the cascaded FWM in our experiments [32].

Initially, 515 nm pulses are produced via second harmonic generation in a beta barium borate (BBO) crystal (Eksma Optics). After second harmonic generation, the pulses centered at 1030 nm and 515 nm are separated to allow for independent adjustment of their temporal delay and polarization state before being recombined with parallel polarization and coupled into the NC-HCF using a 10 cm focal length lens.

Additionally, the fiber mounting system was designed to support operation under both a constant pressure and pressure gradient. The gradient is established by injecting gas at the fiber end where the laser is coupled and allowing the gas to flow through the fiber into vacuum at the other end of the fiber. Xenon was chosen as the active medium due to its high third-order nonlinearity [33].

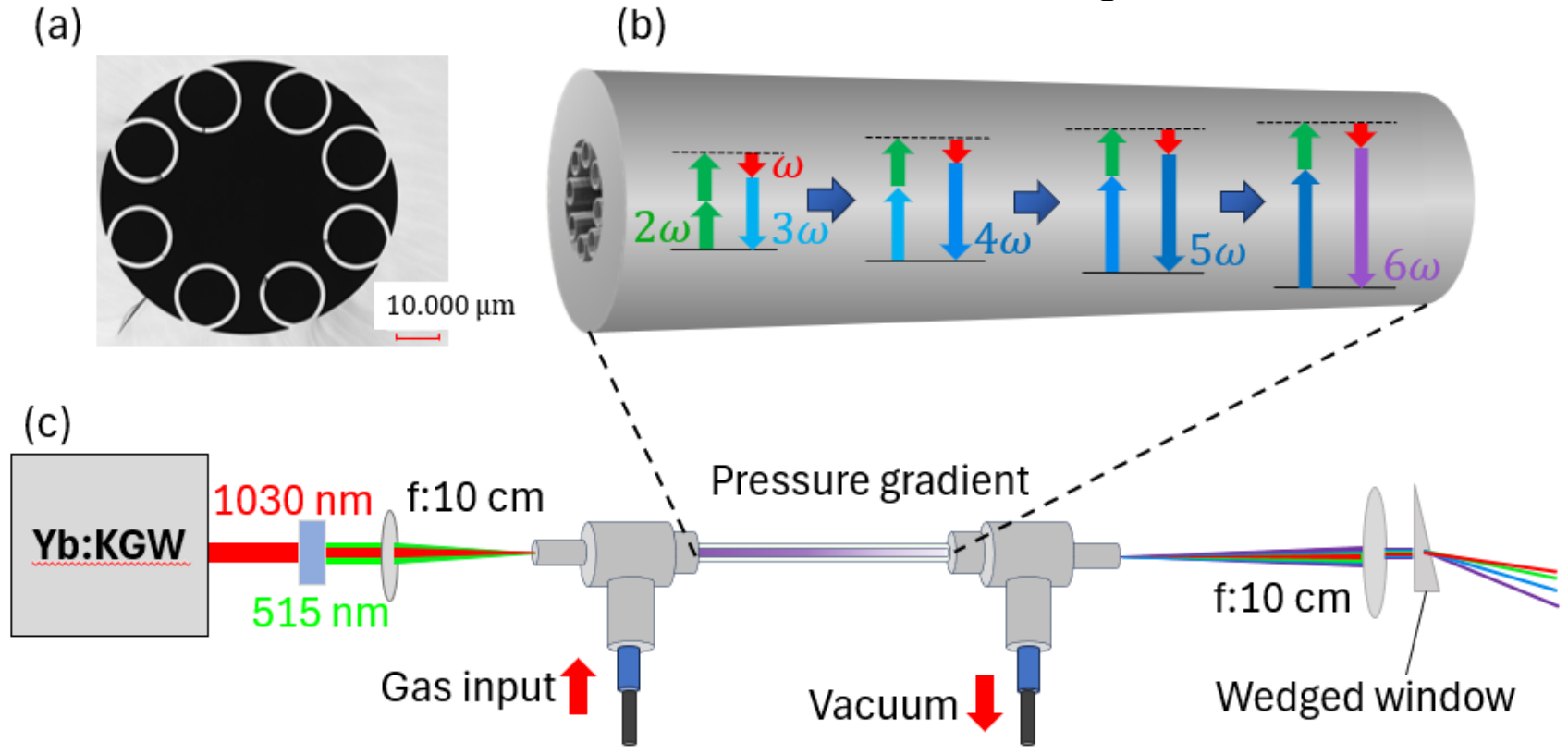

Figure 1: Schematic of the experimental setup and the cascaded harmonic generation process. (a) Microscopic image of the cross section of the NC-HCF. (b) Energy-level diagrams illustrating the cascaded FWM interactions within the NC-HCF, where the fundamental ($\omega$) and second-harmonic ($2\omega$) beams drive the generation of subsequent frequencies up to the sixth harmonic ($6\omega$). (c) Detailed optical layout showing the focusing and collimating lenses (f = 10 cm), the gas input and vacuum connections establishing the longitudinal pressure gradient, and the wedged window used to spatially separate the generated harmonics at the output.

Spectral measurements down to 200 nm were performed using a compact spectrometer (Ocean Optics HR4000). The intensity of the harmonic at 343 nm was measured using a standard silicon photodiode sensor (Thorlabs S120VC), while for the harmonic at 257 nm a photomultiplier was used (Hamamatsu H5783-04) with a lock-in detection. The spectrum of the 6$^{th}$ harmonic falling in the VUV region (172 nm) was recorded by a vacuum VUV spectrometer (McPherson 234/302 monochromator and Andor DO420A CCD camera).

At the output of the fiber, a $BaF_2$ lens with a 10 cm focal length collimated the beams onto a 4° wedged window to separate the harmonics for spectral measurements down to 200 nm. For power measurements, the harmonic at 343 nm was isolated using a dichroic mirror, while the harmonic at 257 nm was directed through a monochromator for better isolation of background scattered light from other wavelengths before its detection with the photomultiplier.

Due to the strong atmospheric absorption of VUV light, for detection below 200 nm the optical path from the fiber output to the detection system was kept under vacuum. In this configuration, the $BaF_2$ lens and the wedged window were replaced by a VUV filter with peak transmission centered at 180 nm (Teledyne Acton Optics 180-B-5D).

## Results and discussion

The coupled-wave equation that describes the evolution of the interacting fields in a nonlinear optical process is given by Eq. (1) [34,35], where $\mathcal{E}_j$ is the slowly varying complex amplitude of the electric field for the $j$-th interacting wave propagating along the $z$ axis, $\mu_0$ is the vacuum permeability, $\omega_j$ is the angular frequency, and $k_j$ is the wavenumber of the $j$-th wave. The term $\mathcal{P}_j$ denotes the complex amplitude of the nonlinear polarization driving the field at frequency $\omega_j$, and $\Delta k_j$ is the phase mismatch between the field and the parametric component of the nonlinear polarization at frequency $\omega_j$, defined by Eq. (2), where $k_j^p$ is the wavenumber of the corresponding nonlinear polarization. One key point for achieving efficient conversion is minimizing the phase mismatch [36]. Under the phase matching condition, $\Delta k_j = 0$, it can be seen from Eq. (1) that the electric field amplitude $\mathcal{E}_j$ grows linearly with the interaction length.

$$\frac{\partial \mathcal{E}_j}{\partial z} = i\frac{\mu_0 \omega_j^2}{2k_j}\,\mathcal{P}_j e^{i\Delta k_j z} \tag{1}$$

$$\Delta k_j = k_j^p - k_j \tag{2}$$

In harmonic generation driven by the FWM process in a gas-filled NC-HCF, the energy conservation for the third harmonic of the fundamental beam is written as $\omega_3 = 2\omega_2 - \omega_1$, with $\omega_2 = 2\omega$ and $\omega_1 = \omega$. In the cascaded harmonic generation through FWM in NC-HCF, as illustrated in Figure 1, the strong field at $\omega_2$ generated externally with an efficient phase matched second harmonic generation in a BBO crystal participates in a sequence of FWM processes where the field generated at the next harmonic frequency contribute in a new FWM leading to generation of the next harmonic and so on.

In this work, our focus is on the optimization and analysis of the initial harmonics generated in the sequence, specifically the third and fourth harmonics. In the cascaded FWM process within a gas-filled NC-HCF, the cascaded begins with the generation of the third harmonic through the energy conservation relation $\omega_3 = 2\omega_2 - \omega_1$, where $\omega_2 = 2\omega$ is the externally generated second harmonic and $\omega_1 = \omega$ is the fundamental frequency. In the cascaded harmonic generation scheme illustrated in Figure 1, the intense second-harmonic field, produced beforehand by an efficient phase-matched second-harmonic generation process, drives a sequence of FWM interactions. In each step, the newly generated harmonic participates in the subsequent FWM processes together with the persistent second-harmonic field, producing the next higher-order harmonic. A key feature of this mechanism is that the strong second-harmonic field is involved in every FWM stage of the cascade harmonic generation, continuously transferring energy to progressively higher harmonics. This repeated participation substantially enhances the conversion efficiency, enabling the generation of coherent radiation extending into the vacuum ultraviolet (VUV) spectral region [30].

As a first step in our analysis, the integration of Eq. (1) from zero to $L$ (length of nonlinear interaction), leads to the well-known equation connecting the intensity of the new frequency generated as function of phase mismatch and the interaction length of the nonlinear interaction:

$$I_j(z=L) = c\frac{\mu_0\omega_j^2}{4}L^2\left|\mathcal{P}_j\right|^2\mathrm{sinc}^2\left(\frac{\Delta k_j L}{2}\right) \tag{3}$$

where for the generation of the $\omega_3$ by FWM, the term $\mathcal{P}_j$ is given by:

$$\mathcal{P}_3 = \epsilon_0\frac{3}{4}\chi^{(3)}E_2^2E_1^* \tag{4}$$

where $\epsilon_0$ is the vacuum permittivity, $\chi^{(3)}$ is the third order susceptibility, $E_2$ and $E_1$ are the amplitudes of the incident fields at frequencies $\omega_2$ and $\omega_1$ respectively. It is important to highlight that Eq. (3) is valid for a constant phase mismatch $\Delta k$ along the interaction length $L$, which is not the condition under which the cascaded FWM for generation of VUV pulses occurs. However, Eq. (3) will be important to validate the model used to describe the phase mismatch term in the NC-HCF filled with Xe, as we will see, will be modeled by the Marcatili and Schmeltzer theory [37].

Despite the nontrivial mechanism that describes the optical confinement in an NC-HCF [38], away from the spectral windows with high attenuation, its propagation can be described by the method developed by Marcatili and Schmeltzer to describe the propagation in dielectric capillaries [37,39]. According to this model, since the refractive index of a gas is proportional to the pressure, $n-1 = p\delta(\lambda)$, where $p$ is the gas pressure, and $\delta(\lambda)$ is the dispersion, for a FWM process to generate $\omega_3$ as discussed, the phase mismatch, defined in Eq. (2), will be given by:

$$\Delta k(p) = 2\pi p\left(2\frac{\delta_2}{\lambda_2}-\frac{\delta_1}{\lambda_1}-\frac{\delta_3}{\lambda_3}\right)-\frac{1}{4\pi a^2}\left(2u_2^2\lambda_2-u_1^2\lambda_1-u_3^2\lambda_3\right) \tag{5}$$

where $u_j$ is a constant associated with wave propagation mode, and $a$ is the hollow waveguide radius.

In Eq. (5), the first term in the right-hand side is pressure dependent, and it may be used to balance the second term in order to obtain phase matching. Since the third order susceptibility $\chi^{(3)}$ is proportional to the number density of the gas, and, therefore, is linearly dependent on the gas pressure [33], Eq. (3) can be written as:

$$I_{\omega_3}(p) \propto p^2\mathrm{sinc}^2\left(\frac{\Delta k(p)L}{2}\right) \tag{6}$$

In cascaded FWM in NC-HCF, each harmonic generated drive, together with the strong second harmonic field, the creation of the next harmonic, as illustrated in Figure 1 and discussed previously. Analyzing the phase mismatch for the FWM cascaded process reveals that the phase-matching pressure decreases with increasing harmonic order as shown in Figure 2. Therefore, the cascaded FWM in NC-HCF requires the application of a pressure gradient along the fiber with the pressure decreasing towards the field's propagation direction. This ensures that the optimal phase-matching pressure for each harmonic is locally satisfied at specific points along the propagation axis.

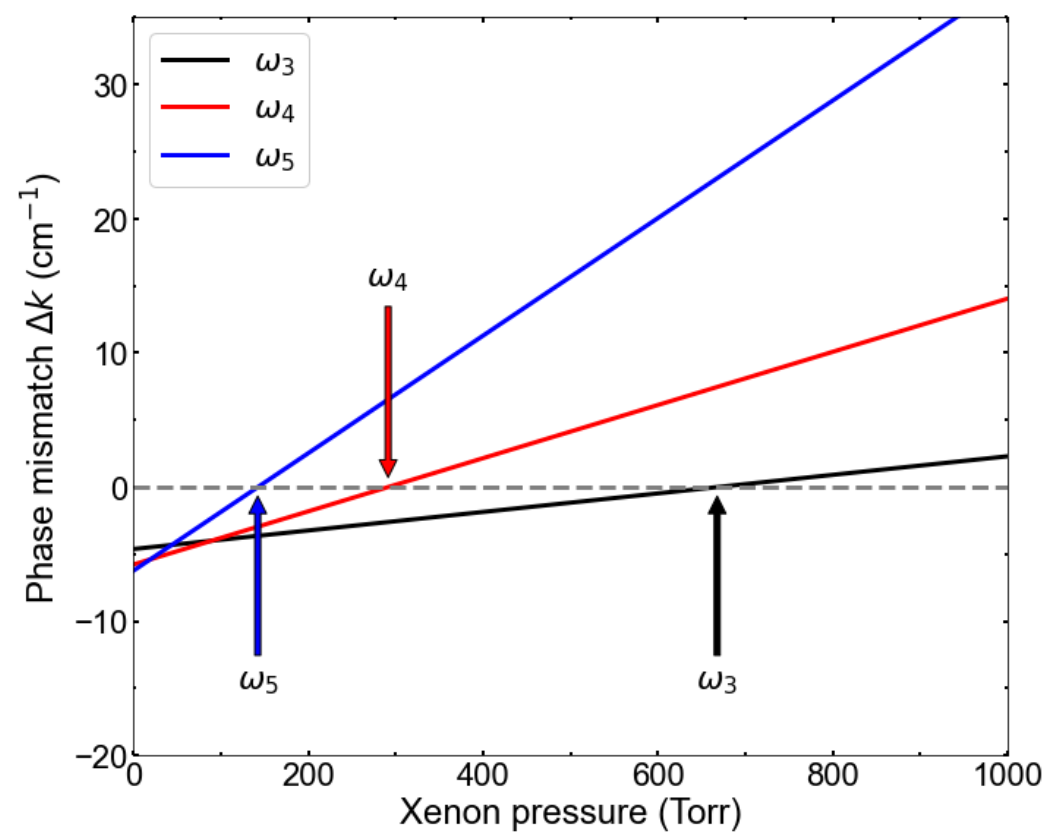

Figure 2: Phase mismatch as a function of xenon pressure for cascaded harmonic generation under constant pressure along the NC-HCF. The black line refers to the process $\omega_3 = 2\omega_2 - \omega_1$, the red to $\omega_4 = \omega_3 + \omega_2 - \omega_1$ and the blue to $\omega_5 = \omega_4 + \omega_2 - \omega_1$.

The pressure profile can be described by [40]:

$$p(z) = \sqrt{p_{in}^2 - \frac{z}{L}\left(p_{in}^2 - p_{out}^2\right)} \tag{7}$$

where $p_{in}$ is the pressure at the fiber input, and $p_{out}$ is the pressure at the output.

Since $\Delta k(p)$ is now $\Delta k[p(z)]$, the accumulated phase difference in Eq. (1) is no longer linear with the propagation distance in the integration that leads to the output field amplitude in the generated harmonic. Therefore, the field generated by FWM under the pressure gradient will be given by:

$$\mathcal{E}_{\omega_j} \propto \int_0^L p(z) \exp\left\{ i \int_0^z \Delta k_j[p(z')] dz' \right\} dz \tag{8}$$

As will be shown throughout the discussion, applying Eq. (6) and (8) to model third-harmonic generation under constant-pressure and pressure-gradient conditions, while considering the phase mismatch given by Eq. (5), which accounts only for the linear refractive index, yields reasonably good agreement with experimental measurements. To further assess the validity of this simplified approach, we compared the experimental results with a more comprehensive model that includes the contributions of the intensity dependent nonlinear refractive index to the phase mismatch. These simulations were implemented using the Julia package Luna.jl [41].

The simulations performed with the Julia package Luna.jl employed parameters corresponding to the experimental conditions: a pulse duration of 170 fs; pulse energy of 1 μJ for the second harmonic and 0.2 μJ for the fundamental frequency; a fiber length of 12 cm; and a fiber core diameter of 35 μm, chosen to obtain a better match between the simulation and the experimental curve, given the uncertainty in this parameter. This is justified since the core diameter can vary between different fabrication batches. The function prop_capilary used in the Luna.jl simulation also employs the Marcatili and Schmeltzer theory for pulse propagation in dielectric capillaries. However, since the propagation losses in an NC-HCF are significantly lower than those in a dielectric capillary, these losses were neglected in our simulation. To compare the semi-analytical model given by Eq. (6) with the Luna.jl simulation, we used the same parameters for the NC-HCF in both methods, yielding the curves in Figure 3-a). It can be seen that, despite their very similar functional form, the two curves are shifted relative to each other, with the curve based in the semi-analytical model peaking in slightly higher pressure (733 torr for Luna.jl simulation and 750 torr for the semi-analytical simulation). This difference can be attributed to intensity-dependent effects that the semi-analytical model does not account. To corroborate it, the pulse energy of both fields was divided by a factor of 100, resulting in the blue dotted curve in Figure 3-b) that fully agrees with the semi analytical model.

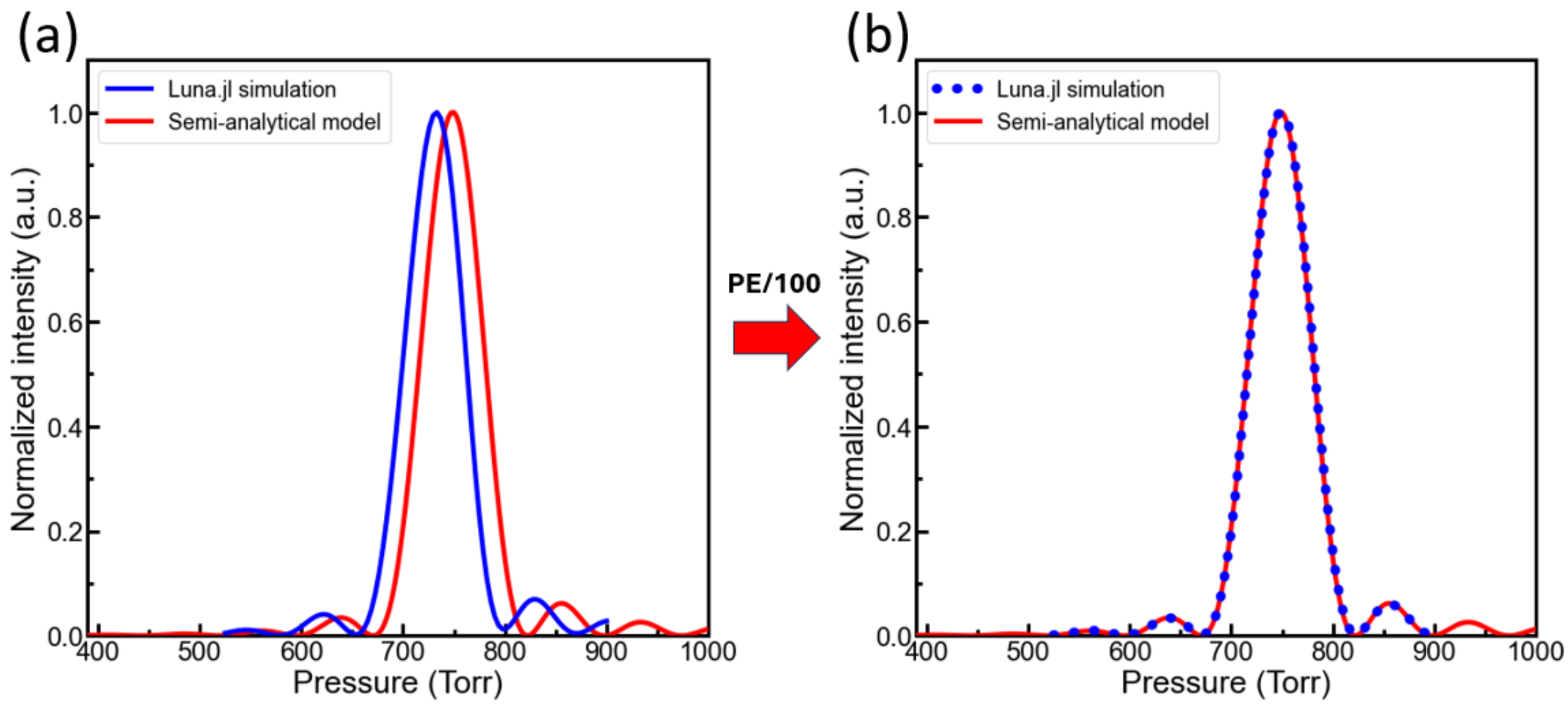

Figure 3: Third-harmonic generation curves via FWM as a function of pressure, obtained from the semi-analytical model (red) and the Luna.jl simulation (blue). (a) Results using a pulse energy of 1 μJ for the $2\omega$ field and 0.2 μJ for $\omega$ (170 fs pulses). (b) Mitigation of the peak position difference by reducing the pulse energy (PE) by a factor of 100. A fiber diameter of 35 μm and a length of 12 cm were used for generating both theoretical curves.

We measured the third harmonic intensity as a function of Xe pressure while maintaining a constant pressure profile along the fiber. The experimental results are shown in Figure 4 together with the simulated pressure dependent curves. The Luna.jl simulation is in excellent agreement with the experimental data, whereas the semi-analytical model predicts a peak at a slightly higher pressure, as already discussed (Figure 3). Now, taking a different approach than the one that leads to Figure 3-b), where the pulse energy was reduced by a factor of 100, we added 10 $rad\,m^{-1}$ to the phase mismatch term ($\Delta k$) on the semi-analytical model. The resulting curve, shown by the solid red line, agrees closely with both the experimental data and the Luna.jl simulation. As discussed in connection with Figure 3, this initial discrepancy can be attributed to intensity-dependent refractive index effects. The additional 10 $rad\,m^{-1}$ contribution to $\Delta k$ likely provides an effective correction for the nonlinear index of refraction dispersion, which modifies the propagation constants of the field with different frequencies involved in the FWM process by different amounts.

With the confirmation that Marcatili and Schmeltzer theory can be used to model the phase mismatch for the FWM process in the NC-HCF used in this work, we moved forward with our study on the optimization of the third and fourth harmonics and their modeling under the condition of pressure gradient aiming to optimize the cascade FWM process for generation of VUV ultrashort pulses under high repetition rate.

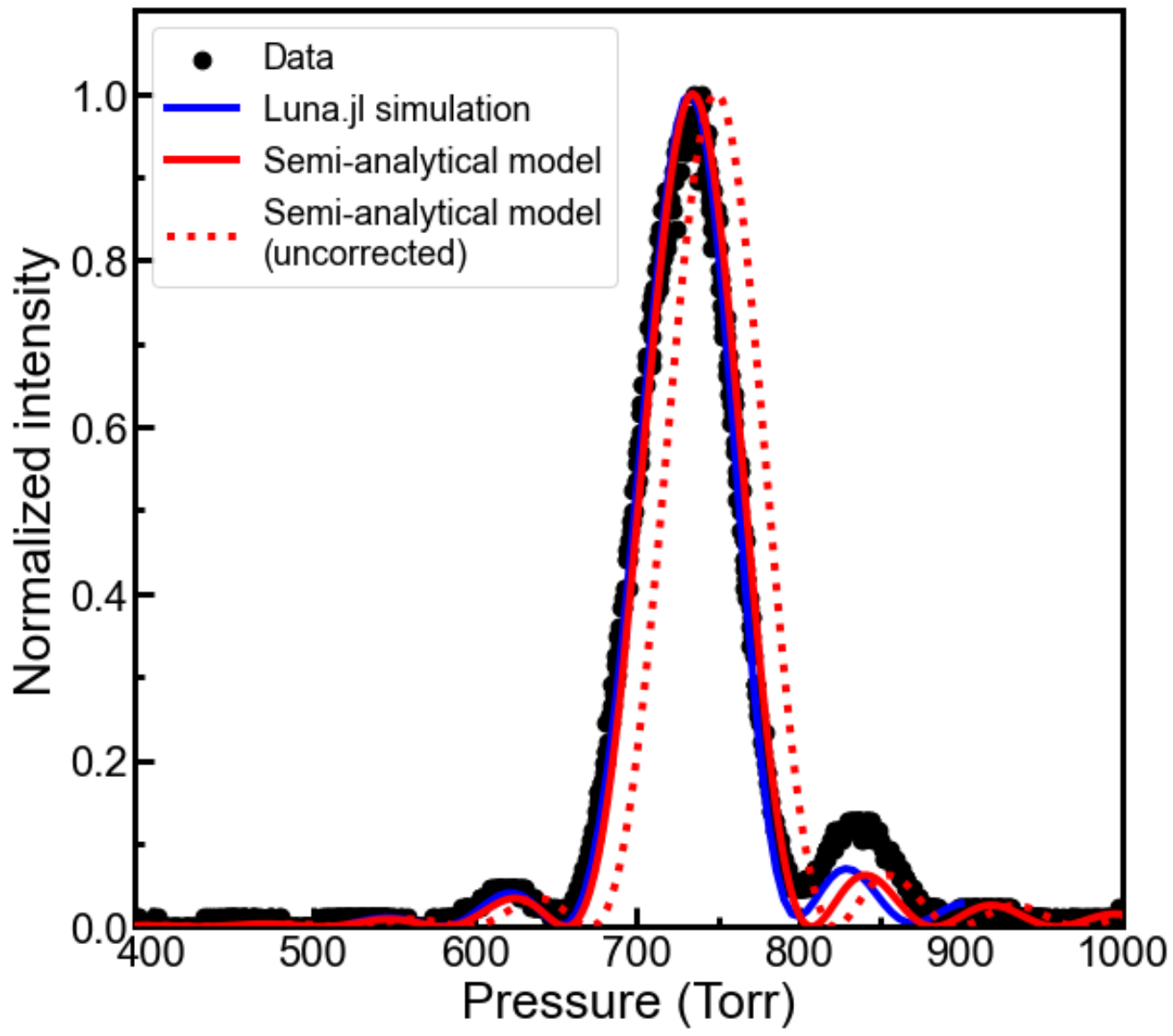

Figure 4: Normalized intensity of the generated third harmonic as a function of uniform gas pressure. The experimental data (black dots) are compared with numerical simulations performed in Luna.jl (solid blue line) and a semi-analytical model. The analytical description is shown both uncorrected (dotted red line) and with an empirical 10 $rad\ m^{-1}$ correction to the phase mismatch $\Delta k$ (solid red line) to account for intensity-dependent nonlinear effects.

To study the third-harmonic generation centered at 343 nm under a pressure gradient, we varied the input gas pressure at the end of the fiber where the laser beams driving the FWM process were coupled while maintaining the output under vacuum. As shown in Figure 5, the $\omega_3$ intensity exhibits a broad peak at ~780 torr, which is slightly higher than the optimum pressure observed for the constant-pressure condition (~730 torr). This result could be expected based on the fact that, in a decreasing pressure gradient, starting at a pressure higher than the optimal constant-pressure value allows the local pressure to gradually decrease along the fiber until perfect phase-matching is reached. This dynamic minimizes the total accumulated phase mismatch over the effective interaction length, maximizing the conversion efficiency. Similar to the constant-pressure configuration, we modeled this process using a semi-analytical approach based on Eq. (8) alongside Eq. (5) and (7), solving the integral numerically. Following this approach, by calculating the third harmonic intensity dependence with input pressure, we obtained the red curve shown in Figure 5 with the phase mismatch due to the nonlinear refractive index already corrected, as discussed for the case of constant pressure. The dependence obtained based on the Luna.jl simulation that takes the nonlinearity of refractive index into account is shown alongside in the blue curve in Figure 5. A notably good agreement is observed between both models and the experimental data for pressures above the optimal pressure up to ~850 torr, where an increasing disagreement for the amplitudes arise although the functional form showing a local minimum at ~ 900 torr and a local maximum at ~ 960 torr are correctly captured by both models.

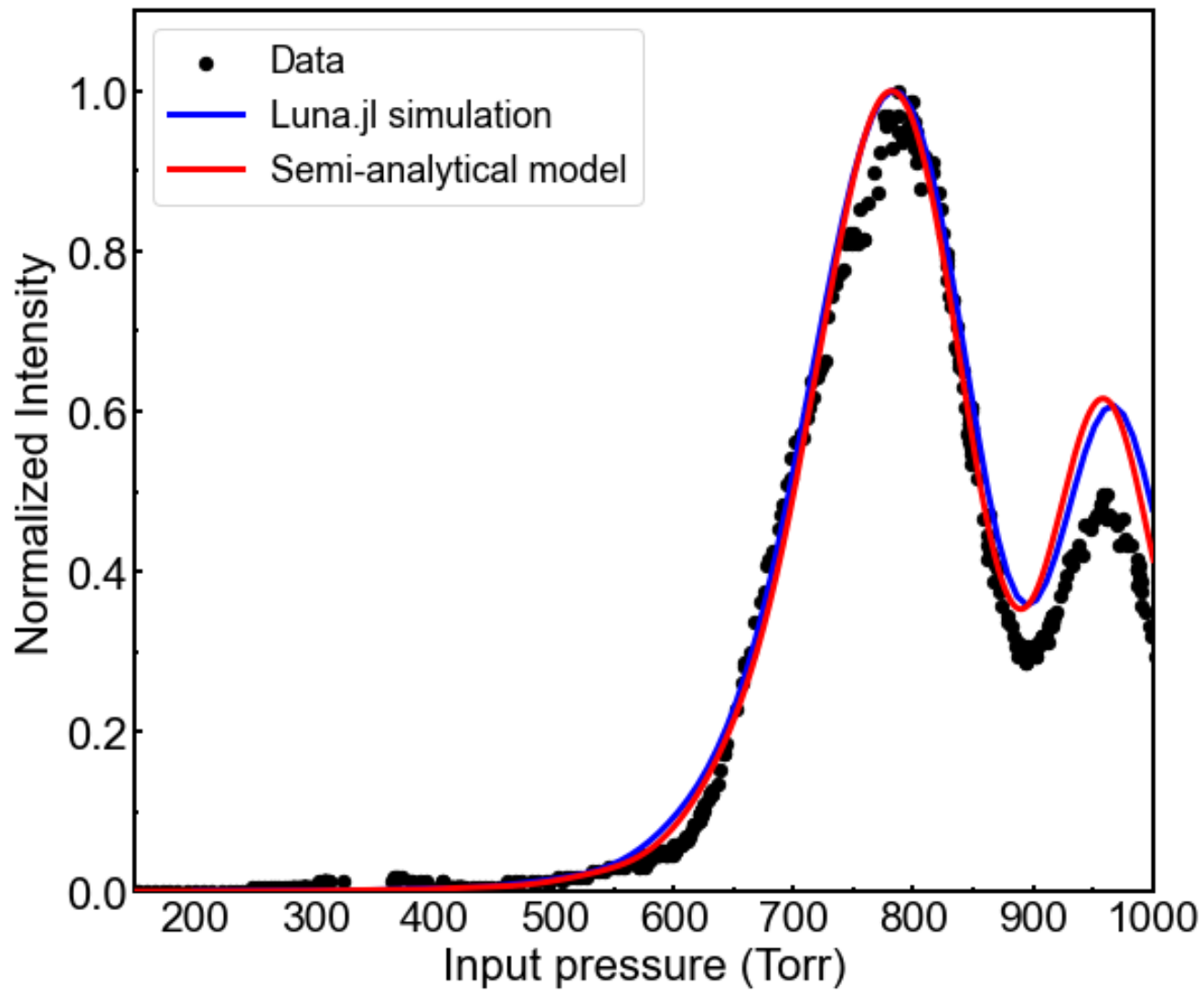

Figure 5: Normalized intensity of the third harmonic as a function of the input gas pressure under a pressure gradient. The experimental data exhibits a well-defined maximum, clearly demonstrating the critical need for precise pressure tuning to optimize the generation process. Both the Luna.jl simulation and the corrected analytical model show good agreement with the experimental results. In this case, the core diameter used for the simulations was 37.1 μm.

These results demonstrate that our numerical model accurately reproduces the third harmonic generation in the Xe filled under pressure gradient pressure, method employed in the cascade harmonic generation up to VUV energies. This confirmation is an essential step, as it validates our modeling of the FWM phase mismatch for the NC-HCF, providing the necessary foundation to predict the optimal conditions required to extend the generation to higher-order harmonics.

Since the developed simple semi-analytical model accurately describes the third harmonic generation at 343 nm, we applied it to the next harmonic in the cascaded process, namely the fourth harmonic centered at 257 nm. Its generation efficiency dependence on the input pressure was experimentally characterized and compared with the predictions based on the same model. We measured both harmonics simultaneously varying the input pressure, and the resulting data are shown in Figure 6. As the harmonic order increases, there are more possible frequency combinations that result in the same harmonic. For example, the fourth harmonic ($\omega_4$) can be generated via: $\omega_4 = \omega_3 + \omega_2 - \omega_1$ and $\omega_4 = 2\omega_3 - \omega_2$. Although, it's important to notice that the second process depends quadratically on the field amplitude at $\omega_3$, which is also generated along the fiber and only achieves an amplitude compared to the second harmonic field towards the end of the fiber while the intense field at $2\omega$ is available since the fiber input. Therefore, we consider only the first process, which depends linearly in the amplitude of the third harmonic field $E_3(z,p)$ generated along the fiber. By determining the fourth harmonic field using the same method as for the third, but now accounting for the variation of $E_3(z,p)$ along the fiber, we obtained the solid blue line in Figure 6. It can be seen that the model satisfactorily describes the pressure dependence of the field at $\omega_4$.

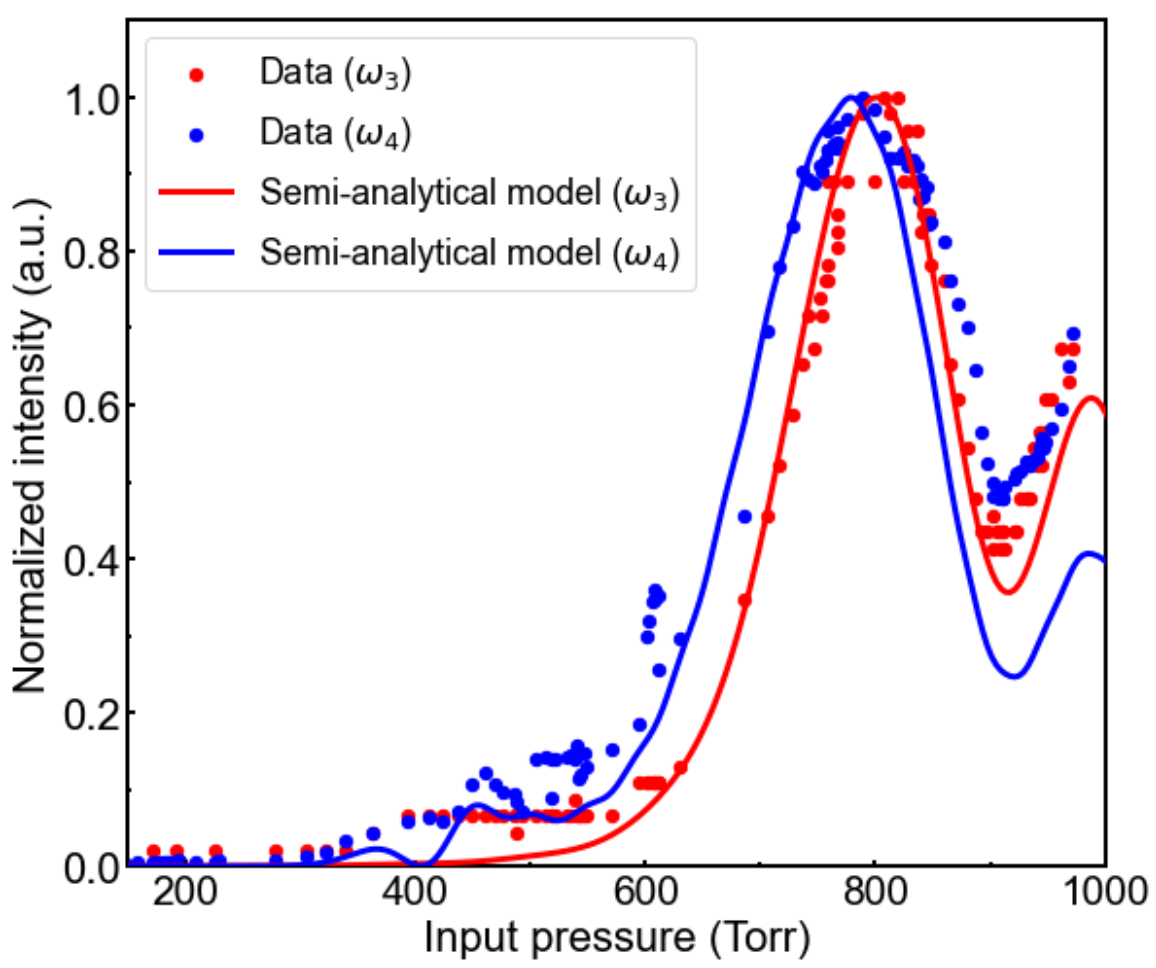

Figure 6: Simultaneous measurement of the generated $\omega_3$ and $\omega_4$ harmonics as a function of gas pressure. Experimental data are shown for $\omega_3$ (red symbols) and $\omega_4$ (blue symbols), together with their respective semi-analytical model predictions (solid lines). The theoretical curve for $\omega_4$ (solid blue line) based on the semi-analytical model accounts for the cascaded generation process driven by the third harmonic field $\omega_3$ field generated along the fiber according with the red solid line curve, showing reasonable agreement with the experimental pressure dependence (core diameter 37.1 μm).

As shown in Figure 6, additionally to the characterization, modelling and optimization of third and fourth harmonics (343 nm and 257 nm) in the conditions that lead to cascaded FWM, we successfully showed that the fourth harmonic resulting from the cascaded FWM is optimized at the same optimum pressure as for the third harmonic field. To confirm the relevance of the results presented in this work for the ultrafast VUV generation using this platform, we demonstrate the capability of our optimized setup by recording the extended cascaded FWM spectrum up to VUV under the pressure gradient conditions that maximize the third harmonic field. Our set-up allowed us to observe the cascaded harmonic generation, specifically up to the sixth harmonic at 172 nm (7.2 eV) as shown in Figure 7. Harmonics with energies higher than 7.2 eV weren´t observed due to the VUV filter available centered at 180 nm as discussed in the Experimental Section. This confirms that the optimized $\omega_3$ field effectively acts as a driving stage for the extended harmonic cascade within the NC-HCF.

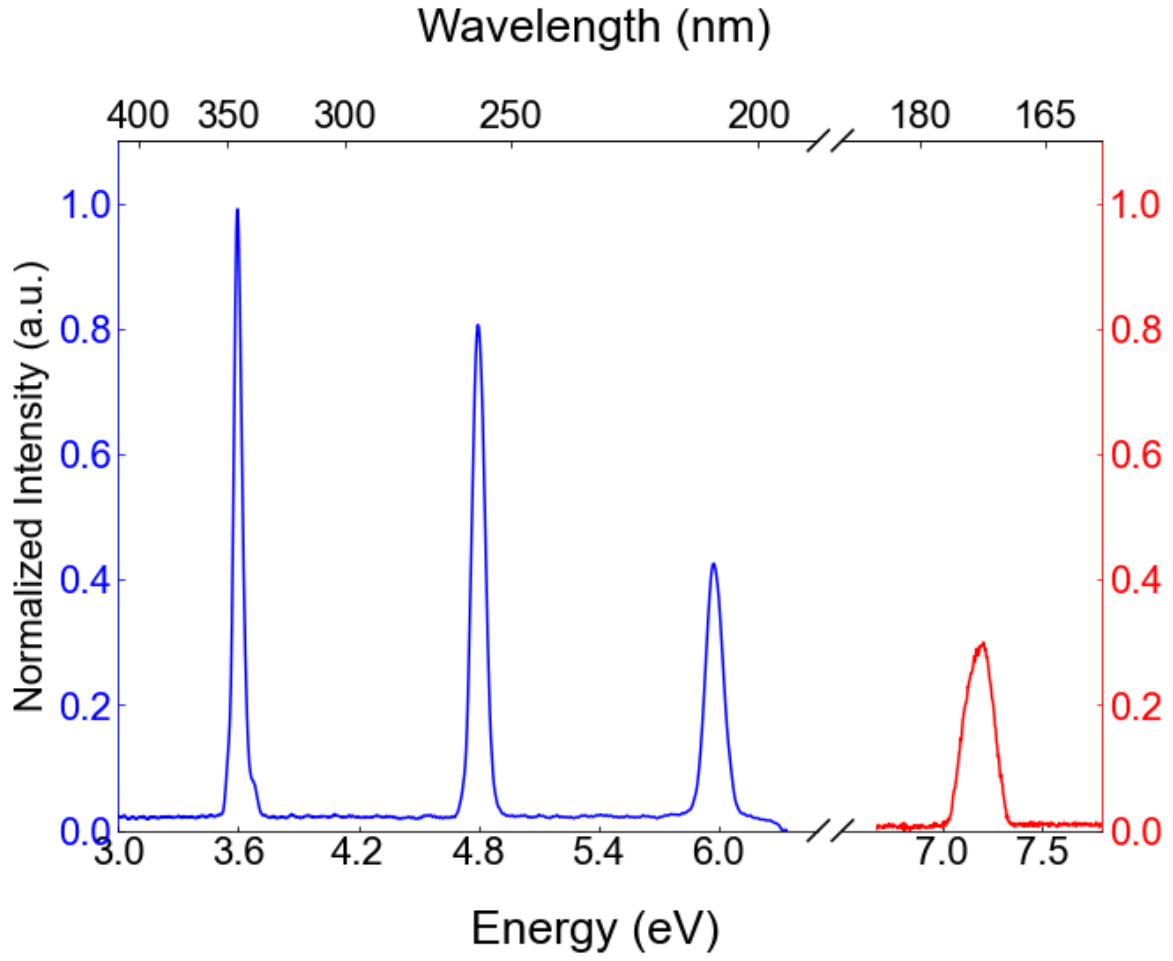

Figure 7: Spectrum of the cascaded harmonics generated. The blue curve (left axis), measured using the Ocean Optics spectrometer, shows the $\omega_3$ (3.6 eV), $\omega_4$ (4.8 eV) and $\omega_5$ (6.0 eV) harmonics. The red curve (red axis), recorded by VUV spectrometer, shows the $\omega_6$ (7.2 eV) harmonic. The signals are plotted on independent axes because the relative intensity between the two detectors is uncalibrated.

## Conclusion

In this work, we thoroughly characterized cascaded FWM processes in a Xe-filled NC-HCF focusing on the optimization of the third and fourth harmonics with support from both semi-analytical and numerical simulations. The semi-analytical model successfully described FWM-based harmonic generation under the longitudinal pressure gradient required to extend the cascade into the VUV. Its predictions agreed well with both the experimental results and numerical simulations performed using the Luna.jl package. We identified the optimum input pressure for maximizing the third- and fourth-harmonic fields under pressure-gradient conditions and showed that the two harmonics reach their maxima at approximately the same input pressure. Using this optimum pressure, we extended the cascaded FWM process into the VUV, reaching the sixth harmonic at 172 nm, with the measured spectral range limited by the available VUV filter. This detailed characterization of the early stages of cascaded FWM in Xe-filled NC-HCFs provides deeper insight into the interplay among phase matching, pressure-gradient-driven dispersion, and harmonic buildup throughout the cascaded FWM, while also offering practical guidance for designing high-repetition-rate ultrafast VUV sources for steady state or time-resolved photoemission spectroscopies and microscopies.

## Acknowledgment

The authors thank professors Lino Misoguti and Flavio Caldas da Cruz for sharing equipment. This work was supported by Fundação de Amparo à Pesquisa do Estado de São Paulo (2022/03035-9, 2024/00369-9); Fundação de Amparo à Pesquisa do Estado de Minas Gerais (RED-00046-23, APQ-00197-24, APQ-01618-25); Conselho Nacional de Desenvolvimento Científico e Tecnológico (305024/2023-0, 402723/2024-4, 409174/2024).